\documentstyle[12pt]{article}
\begin{document}
\tolerance=5000
\def\be{\begin{equation}}
\def\ee{\end{equation}}
\def\bea{\begin{eqnarray}}
\def\eea{\end{eqnarray}}
\def\nn{\nonumber \\}
\def\cF{{\cal F}}
\def\det{{\rm det\,}}
\def\Tr{{\rm Tr\,}}
\def\e{{\rm e}}
\def\etal{{\it et al.}}
\def\erp2{{\rm e}^{2\rho}}
\def\erm2{{\rm e}^{-2\rho}}
\def\er4{{\rm e}^{4\rho}}
\def\etal{{\it et al.}}
\def\dv{\left( 1 + {a \over \lambda}\e^{\lambda\sigma^-}
\right)}
\def\ds{\left( 1 + {a \over \lambda}
\e^{\lambda(\sigma^- - \sigma^+ + \sigma_0)}
\right)}

\  \hfill 
\begin{minipage}{3.5cm}
NDA-FP-35 \\
June 1997 \\
\end{minipage}

\ 

\vfill

\begin{center}

{\Large\bf Trace anomaly induced effective action 
for 2D and 4D dilaton coupled scalars}

\vfill

{\large\sc Shin'ichi NOJIRI}\footnote{
e-mail : nojiri@cc.nda.ac.jp}
and
{\large\sc Sergei D. ODINTSOV$^{\spadesuit}$}\footnote{
e-mail : odintsov@quantum.univalle.edu.co, \\
odintsov@kakuri2-pc.phys.sci.hiroshima-u.ac.jp}

\vfill

{\large\sl Department of Mathematics and Physics \\
National Defence Academy \\
Hashirimizu Yokosuka 239, JAPAN}

{\large\sl $\spadesuit$ 
Tomsk Pedagogical University \\
634041 Tomsk, RUSSIA \\
and \\
Dep.de Fisica \\
Universidad del Valle \\
AA25360, Cali, COLOMBIA \\
}

\vfill

{\bf ABSTRACT}

\end{center}

The spherically symmetric reduction of higher dimensional 
Einstein-scalar theory leads to lower dimensional dilatonic 
gravity with dilaton coupled scalar (for example, from 4D to 
2D system). We calculate trace anomaly and anomaly induced 
effective action for 2D and 4D dilaton coupled scalars. 
The large-N effective action for 2D quantum dilaton-scalar 
gravity is also found. These 2D results maybe applied for 
analysis of 4D spherical collapse. The role of new, 
dilaton dependent terms in trace anomaly for 2D black 
holes and Hawking radiation is investigated in some 
specific models of dilatonic gravity which represent 
modification of CGHS model.
The conformal sector for 4D dilatonic gravity is constructed. 
Quantum back-reaction of dilaton coupled matter is briefly discussed 
(it may lead to the inflationary Universe with non-trivial dilaton).


\newpage

\section{Introduction}

There are few motivations to study 2D dilatonic 
gravity models. 
First of all, it is often easier to study 2D models 
than their 4D analogs.
Especially, it may happen on quantum level.
For example, for renormalizable classically solvable dilaton gravity coupled with minimal scalar matter, 
the problem of Hawking radiation and 
the back-reaction of 
matter to 2D black hole maybe well-understood in 
$1/N$ expansion \cite{12}.
Modifications of CGHS model and Hawking radiation 
there have been investigated in refs.\cite{RST,A,BPP} 
and many other works (for a review, see \cite{Rev}).
Second, some 2D dilatonic gravities are string inspired 
ones.
They may serve as a laboratory for better understanding 
of string theory itself.
Third, if one starts from the 4D Einstein-scalar 
or 4D Einstein-Maxwell-scalar theory then using a 
sherically symmetric reduction anzatz \cite{8}, 
one obtains the action for one of the 2D dilatonic gravity 
models with scalars.
For example, applying such anzatz (\ref{12}) to 4D 
Einstein-Maxwell-minimal scalar theory and 
integrating of the angular modes, one gets
\bea
\label{0}
S&=&-{1 \over 16\pi G}\int d^2 x \sqrt{-g} \e^{-2\phi}
\left( R+2(\nabla \phi)^2 + 2\e^{2\phi}-2Q^2 \e^{4\phi}
\right) \nn
&& + {1 \over 2}\int d^2 x \sqrt{-g} \e^{-2\phi}
(\nabla \chi)^2\ .
\eea
Hence, 4D spherically symmetric collapse maybe 
understood in terms of 2D dilatonic gravity.

In difference with CGHS model and its modifications, 
we have the scalar field non-minimally coupled 
with dilaton.
Then generalization of CGHS model and study of the 
Hawking radiation \cite{H} 
 in generalized model in large-$N$ 
approximation (then one has to consider $N$ scalars 
in above model) requires the calculation of the 
trace anomaly for dilaton coupled scalar.

Such trace anomaly has been recently found for above 
model in ref.\cite{5} and in case of an arbitrary 
dilaton-scalar coupling function $f(\phi)$ in 
\cite{4}. 
The correspondent trace anomaly induced effective action 
has been also calculated \cite{5,4}.
(Actually, the trace anomaly is proportional to $b_2$-coefficient of
 Schwinger-De Witt expansion which for general dilaton coupled scalar 
has been found some time ago in ref.\cite{6}).

The natural next step is to discuss the quantum gravity contributions 
to such action and its applications to 2D black holes and Hawking 
radiation. The question of 4D generalization is also of interest.

The present paper is devoted to the study of this circle of questions. 
In the next section we discuss general model of dilatonic 
gravity \cite{1,2}.
The trace anomaly and anomaly induced effective action are calculated 
for $N$ dilaton coupled two-dimensional scalars in general 
form and in large-$N$ limit. (The contribution of quantum dilaton is also 
included). In the case when dilaton-scalar gravity is quantized the 
one-loop divergent effective action is used to find the large-$N$ non-local 
finite action. This action almost coincides with anomaly induced large-$N$ 
action as it should be. 

Section 3 is devoted to the investigation of 2D black holes in modified 
CGHS model (we consider dilaton coupled scalars). The new, dilaton dependent 
terms in trace anomaly make the problem much more complicated than for minimal 
case. In particular, the theory is not classically solvable anymore. 
The new black hole solution for purely induced theory is found. 
For some known 
cases black holes, Hawking radiation and black hole entropy are 
briefly discussed.

In section 4 we calculate trace anomaly and induced effective action for 
dilaton coupled 4D scalar. The motivation to do so is similar to 2D case. 
Let us start from higher dimensional Einstein-scalar theory. After 
spherically symmetric reduction anzats one is left with lower dimensional 
(say 4D) dilatonic gravity with dilaton coupled scalar. 

Section 5 is devoted to the formulation of the conformal sector 
for 4D dilatonic gravity. The classical solutions of such theory 
describe quantum cosmology (with back-reaction of matter). One of the 
solutions for purely induced theory may correspond to 
the inflationary Universe.
In the conclusion we give summary and list of the problems for future research.

\section{One-loop effective action in the large-$N$ 
approximation}

We will start from the dilaton gravity of most 
general form \cite{1,2} interacting with 
scalar matter:
\bea
\label{1}
S&=&-\int d^2x \sqrt g \Bigl\{ {1 \over 2}Z(\phi) 
g^{\mu\nu}\partial_\mu\phi\partial_\nu\phi \nn
&&+C(\phi)R + V(\phi,\chi)
- - {1 \over 2}
f(\phi) g^{\mu\nu}\sum_{i=1}^N\partial_\mu\chi_i
\partial_\nu\chi_i
\Bigr\}
\eea
It includes a dilaton field $\phi$, $N$ real dilaton coupled scalars 
$\chi_i$ and dilaton-dependent couplings $Z$, $C$, $f$. 
The potential $V$ is function of $\phi$ and 
$\chi_i$.

First of all we will be interesting in the study of 
the one-loop divergent effective action for the above 
theory.
We consider two different cases.
Let in the theory (\ref{1}) dilaton $\phi$ and 
scalars $\chi_i$ are the quantum fields, while the 
gravitational field is the external field.
Then one can apply the background field method 
\cite{3} and to calculate the one-loop effective 
action \cite{6} (see Eq.(41) of ref.\cite{6}):
\bea
\label{2}
\Gamma_{{\rm div}}
&=&- {1 \over 2\epsilon}\int d^2x \sqrt g 
\Bigl\{ \left({C'' \over Z}-{N+1 \over 6}\right)R 
+ {V'' \over Z}- {N \over f}{\partial^2 V \over 
\partial \chi^2} \nn
&& + \left({{f'}^2 \over 2fZ}-{f'' \over 2Z}\right) 
(\nabla^\lambda \chi_i)(\nabla_\lambda \chi_i) \nn
&& + \left({N f'' \over 2f}-{N{f'}^2 \over 4f^2} 
- - {{Z'}^2 \over 4 Z^2} \right)
(\nabla^\lambda \phi)(\nabla_\lambda \phi) 
+\left({Nf' \over 2f}-{Z' \over 2Z}\right)
\Delta \phi \Bigr\}
\eea
where $\epsilon=2\pi (n-2)$ and we use the dimensional 
regularization.

The remarkable fact about the system (\ref{1}) 
with $C=V=0$ and the gravitational field being the 
classical one is that the system is conformally 
invariant system. Then on the quantum level  the conformal (or trace) anomaly 
$T$ is given by
\be
\label{3}
\Gamma_{\rm div}={1 \over n-2}\int d^2x \sqrt g b_2\ \, 
\hskip 1cm T=b_2
\ee
>From here one gets (see also \cite{4})
\bea
\label{4}
T&=&{1 \over 24\pi}\Bigl\{ (N+1)R
- -3\left({{f'}^2 \over 2fZ}-{f'' \over 2Z}\right) 
(\nabla^\lambda \chi_i)(\nabla_\lambda \chi_i) \nn
&& - 3\left({N f'' \over f}-{N{f'}^2 \over 2f^2} 
- - {{Z'}^2 \over 2 Z^2} \right)
(\nabla^\lambda \phi)(\nabla_\lambda \phi) 
- - 3\left({Nf' \over f}-{Z' \over Z}\right)
\Delta \phi \Bigr\}
\eea
while for purely scalar field (dilaton is classical) 
all terms with $Z$ in (\ref{4}) disappear \cite{4}.
For a special case $N=1$, $f(\phi)=\e^{-2\phi}$ 
and no quantum dilaton
\be
\label{5}
T={1 \over 24\pi}\left\{ R 
- -6(\nabla^\lambda \phi)(\nabla_\lambda \phi) 
+ 6\Delta \phi \right\}
\ee
That trace anomaly has been recently calculated in 
ref.\cite{5} using zeta-regularization method.
The coefficient of third term in (\ref{5}) 
disagrees with the result of ref.\cite{5}.
The reasons of this disagreement have been discussed in 
ref.\cite{4} (we are using not zeta-regularization 
but dimensional regularization).

Making the conformal transformation of 
metric $g_{\mu\nu}\rightarrow \e^{2\sigma}g_{\mu\nu}$ 
in the trace anomaly, 
using relation
\be
\label{6}
T={1 \over \sqrt g}{\delta \over \delta \sigma}W[\sigma]
\ee
one can find anomaly induced effective action $W[\sigma]$.
In the covariant, non-local form it maybe presented 
as following \cite{4}:
\bea
\label{7}
W&=&-{1 \over 2}\int d^2x \sqrt g \Bigl[ 
{c \over 2}R{1 \over \Delta}R + F_1(\phi)
(\nabla^\lambda \chi_i)(\nabla_\lambda \chi_i) 
{1 \over \Delta}R \nn
&& + \left(F_2(\phi)- {\partial F_3(\phi) \over 
\partial \phi}\right)\nabla^\lambda \phi
\nabla_\lambda \phi {1 \over \Delta}R 
+ R\int F_3(\phi) d\phi \Bigr]
\eea
where
\bea
\label{8}
&& c={N+1 \over 24\pi}\ ,\ \ 
F_1(\phi)=-{1 \over 8\pi}
\left({{f'}^2 \over fZ}-{f'' \over Z}\right), \nn 
&& F_2(\phi)=-{1 \over 8\pi}
\left({N f'' \over f}-{N{f'}^2 \over 2f^2} 
- - {{Z'}^2 \over 2 Z^2} \right)\ ,  \nn
&& F_3(\phi)=-{1 \over 8\pi}
\left({Nf' \over f}-{Z' \over Z}\right)\ .
\eea
Note that for $f=\e^{-2\phi}$, $Z=0$ (i.e., one 
has to omit all $Z$-dependent terms in (\ref{8})) 
the effective action (\ref{7}) has been 
calculated in ref.\cite{5}. Note also that 
actually it is very easy to get large-$N$ limit of 
the effective action (\ref{7}). 
To do so one only has to omit $Z$-dependent terms 
in $F_2(\phi)$, $F_3(\phi)$ and second term in $c$. 
In addition if dilaton is purely classical one should put  
 $F_1=0$.

Let us consider now the theory with the action (\ref{1}) 
as quantum dilaton-matter gravity where all fields: 
$g_{\mu\nu}$, $\phi$ and $\chi_i$ are quantized ones. 
Using the background field method \cite{3}: 
$g_{\mu\nu}\rightarrow g_{\mu\nu} + h_{\mu\nu}$, 
$\phi \rightarrow \phi + \varphi$ where $h_{\mu\nu}$, 
$\varphi$ are quantum fields, and the minimal gauge 
\cite{1}
\be
\label{9}
S_{\rm gf}=-{1 \over 2}\int C_{\mu\nu}\chi^\mu\chi^\nu
\ee
where $\chi^\mu = -\nabla_\nu \bar h^{\mu\nu} + 
{C' \over C}\nabla^\mu \varphi$, $C_{\mu\nu}=-
C\sqrt g g_{\mu\nu}$, $\bar h_{\mu\nu}=h_{\mu\nu}
- -{1 \over 2}g_{\mu\nu}h$, the one-loop effective 
action maybe found.
For pure dilaton gravity it has been obtained in 
ref.\cite{1} and later in refs.\cite{6,7}.

For dilaton-matter gravity with the classical 
action (\ref{1}) the complete result has been 
obtained in ref.\cite{6} (see Eqs.(31), (32)):
\bea
\label{10}
\Gamma_{{\rm div}}
&=&- {1 \over 2\epsilon}\int d^2x \sqrt g 
\Bigl\{ {24 -N \over 6}R + {2 \over C}V 
+ {2 \over C'}V' - {V_{,ii} \over f} \nn
&& + \left({C'' \over C} - {3{C'}^2 \over C^2} 
- - {C'' Z \over {C'}^2} +{N f'' \over 2f}
- -{N{f'}^2 \over 4f^2} \right)
(\nabla^\lambda \phi)(\nabla_\lambda \phi) \nn
&& +\left({C' \over C} - {Z \over C'}
+{Nf' \over 2f}\right)
\Delta \phi \Bigr\}
\eea
Using Eq.(\ref{10}) one can find the one-loop effective 
action for any specific model.

For example, let us take
\bea
\label{11}
&& Z(\phi)=4\e^{-2\phi}\ , \ \ C(\phi)=\e^{-2\phi}\ , \nn
&& V(\phi,\chi)=2\ ,\ \ f(\phi)=\e^{-2\phi}
\eea
in the action (\ref{1}). 
Then the theory (\ref{1}) with dilatonic couplings 
(\ref{11}) could be obtained (for $N=1$) by using 
a spherically symmetric reduction anzatz \cite{8}:
\be
\label{12}
ds^2=g_{\mu\nu}dx^\mu dx^\nu + \e^{-2\phi}d\Omega^2
\ee
from the 4D Einstein-scalar or 4D Einstein-Maxwell-scalar theory 
(in the last case $V=2-2Q^2\e^{2\phi}$). 
Hence, in such case the action (\ref{1}) with 
dilatonic couplings (\ref{11}) may describe the 
radial modes of the extremal dilatonic black holes 
in four dimensions \cite{9}. In other words, 2D dilatonic
black holes may also describe 4D spherically symmetric collapse.

Using general expression (\ref{10}) we may write the 
one-loop effective action for the theory (\ref{11}) 
(keeping $N$ to be an arbitrary integer):
\bea
\label{13}
\Gamma_{{\rm div}}
&=&- {1 \over 2\epsilon}\int d^2x \sqrt g 
\Bigl\{ {24 -N \over 6}R + 4\e^{2\phi} \nn
&& \hskip 2cm
+(N-12)(\nabla^\lambda \phi)(\nabla_\lambda \phi) 
- -N\Delta \phi \Bigr\}
\eea
The theory is one-loop renormalizable one.
Note that recently the very interesting attempt to 
calculate the one-loop effective action (including the local and  
non-local finite terms) for the model (\ref{11})
 has been done in ref.\cite{10}. 
Unfortunately, the result \cite{10} includes a number 
of mistakes. 
In particular, the divergent part of the one-loop 
effective action in the same minimal gauge (\ref{9}) 
of ref.\cite{1} disagrees with the expression (\ref{13}) 
which coincides (at least, in gravitational sector) 
with the independent results of refs.\cite{1,6,7} 
in all cases where such comparison maybe done. 
Hence, the result of ref.\cite{10} contradicts to 
those of refs.\cite{1,6,7} and does not have correct
on-shell limit \cite{11}. One of the reasons of this 
mistake is that for the calculation of the effective action
 in the model (\ref{11}) another dilatonic gravity 
classically equivalent to (\ref{11}) (after 
conformal transformation and dilaton rescaling) 
is used.
However, it was proved in ref.\cite{11} (with 
explicit example) that classically equivalent 2D 
dilatonic gravities (in a sense of conformal 
transformation) are not quantum equivalent 
off-shell.
They lead to different divergent one-loop 
effective actions which coincide only on-shell.

Let us turn  again to the general model 
(\ref{1}). 
In the large-$N$ limit from (\ref{10}) we get
\bea
\label{14}
\Gamma_{{\rm div}}
&=&- {1 \over 2\epsilon}\int d^2x \sqrt g 
\Bigl\{ -{N \over 6}R 
- - {N \over f}{\partial^2 V 
\over \partial\chi\partial\chi} \nn
&& + \left({N f'' \over 2f}
- -{N{f'}^2 \over 4f^2} \right)
(\nabla^\lambda \phi)(\nabla_\lambda \phi) 
+{Nf' \over 2f}\Delta \phi \Bigr\}
\eea
Actually, the expression (\ref{14}) is given by matter,  
matter-graviton and matter-dilaton loops.
It maybe considered as the source for the effective 
trace anomaly, like in (\ref{4}). Integrating such trace anomaly over $\sigma$ in the same way as in 
Eq.(\ref{6}), we will get 
\bea
\label{15}
W&=&-{N \over 2\pi}\int d^2x \sqrt{-g} \Bigl[ 
{1 \over 48}R{1 \over \Delta}R 
- -{1 \over 8}\ln f R \nn
&& -{1 \over 16\pi}{{f'}^2 \over f^2}
(\nabla^\lambda \phi)
(\nabla_\lambda \phi) {1 \over \Delta}R 
+{1 \over 2Nf}\sum_{i=1}^N{\partial^2 V 
\over \partial\chi_i\partial\chi_i}
\e^{{1 \over \Delta}R} \Bigr]\ .
\eea
The expression (\ref{15}) gives the large-$N$ limit 
of the effective action in quantum dilatonic gravity 
(\ref{1}). 
Note the appearence of new non-local term related with 
the scalar potential $V$ (if it presents in the theory). 
Notice that $V$ breaks the conformal invariance 
of the scalar field. That is why it should not be included 
to the system (\ref{1}) when only scalars are quantized. 
Then few more terms of the same structure as the last one 
in (\ref{15}) may be expected in the strict calculation 
of the one-loop finite effective action in dilaton-scalar 
gravity.

The action (\ref{7}) should be used to take into 
account the back-reaction of quantum dilaton-matter 
system to classical dilatonic gravity. 
On the same time the action (\ref{15}) should be 
added to the classical dilatonic matter-gravity action 
if one would like to take into account the back-reaction 
of quantum dilaton-matter gravity (in large-N limit). 
The non-local actions (\ref{7}), (\ref{15}) open the 
way to new generalizations of models like CGHS-model 
\cite{12} where one can find new black hole solutions 
and (or) new terms in the Hawking radiation.
In the next section, we are going to discuss some 
simple properties of above effective actions 
in connection with $2D$ black holes.

\section{2D black holes and Hawking radiation}

We start with the system where the dilaton gravity 
of special form 
\cite{12} couples with the dilaton coupled scalar fields:
\bea
\label{clac}
S_0&=&{1 \over 2\pi}\int d^2x \sqrt{-g} \nn
&&\times\left\{\e^{-2\phi}\left[
R+4 g^{\mu\nu}\partial_\mu\phi\partial_\nu\phi 
+ 4\lambda^2\right]
- - {1 \over 2}f(\phi)\sum_{i=1}^N
g^{\mu\nu}\partial_\mu\chi_i
\partial_\nu\chi_i\right\} \ .
\eea
We would like to consider the modifications of CGHS model
due to back-reaction of dilaton coupled scalars. Hence, we calculate the effective action (\ref{7}) for dilaton coupled 
scalars in large-N approximation (gravitational field  
 is classical field):
\bea
\label{qc}
W&=&-{1 \over 2}\int d^2x \sqrt{-g} \Bigl[ 
{N \over 48\pi}R{1 \over \Delta}R 
- -{1 \over 8\pi}\left({{f'}^2 \over f} - f''\right)
(\nabla^\lambda \chi_i)(\nabla_\lambda \chi_i) 
{1 \over \Delta}R \nn
&& -{N \over 16\pi}{{f'}^2 \over f^2}
\nabla^\lambda \phi
\nabla_\lambda \phi {1 \over \Delta}R 
- -{N \over 8\pi}\ln f R \Bigr]\ .
\eea
 Hence the complete action of our theory 
is given by
\be
\label{ac}
S=S_0 + W
\ee
We treat this theory as a classical system. The background 
scalar field is considered to be zero so we omit all scalar
 terms in above expression.

In the conformal gauge
\be
\label{cg}
g_{\pm\mp}=-{1 \over 2}\e^{2\rho}\ ,\ \ 
g_{\pm\pm}=0
\ee
the equations of motion are obtained by the variation 
over $g^{\pm\pm}$, $g^{\pm\mp}$ and $\phi$  
\bea
\label{eqnpp}
0&=&T_{\pm\pm} \nn
&=&\e^{-2\phi}\left(4\partial_\pm \rho
\partial_\pm\phi - 2 \left(\partial_\pm\phi\right)^2 
\right) \nn
&& +{N \over 12}\left( \partial_\pm^2 \rho 
- - \partial_\pm\rho \partial_\pm\rho \right) \nn
&& +{N \over 8} \left\{
\left( 
\partial_\pm\tilde\phi \partial_\pm\tilde\phi \right)
\rho+{1 \over 2}{\partial_\pm \over \partial_\mp}
\left( \partial_\pm\tilde\phi 
\partial_\mp\tilde\phi \right)\right\} \nn
&& +{N \over 8}\left\{ 
- -2 \partial_\pm \rho \partial_\pm \tilde\phi 
+\partial_\pm^2 \tilde\phi \right\} + t^\pm(x^\pm) \\
\label{req}
0&=&T_{\pm\mp} \nn
&=&\e^{-2\phi}\left(2\partial_+
\partial_- \phi -4 \partial_+\phi\partial_-\phi 
- - \lambda^2 \e^{2\rho}\right) \nn
&& -{N \over 12}\partial_+\partial_- \rho
- -{N \over 8}\partial_+ \tilde\phi 
\partial_- \tilde \phi 
- -{N \over 4}\partial_+\partial_-\tilde\phi \\
\label{eqtp}
0&=& \e^{-2\phi}\left(-4\partial_+
\partial_- \phi +4 \partial_+\phi\partial_-\phi 
+2\partial_+ \partial_- \rho
+ \lambda^2 \e^{2\rho}\right) \nn
&& -{Nf' \over f}\left\{
{1 \over 16}\partial_+(\rho \partial_-\tilde\phi)
+{1 \over 16}\partial_-(\rho \partial_+\tilde\phi)
- -{1 \over 8}\partial_+\partial_-\rho \right\} \ .
\eea
Here 
\be
\label{tilphi}
\tilde\phi = \ln f
\ee
and $t(x^\pm)$ is a function which is determined by 
the boundary condition.

First we consider the large-$N$ limit, where 
classical part can be ignored. Then field equations 
are becoming simpler  
\bea
\label{eqnpp2}
0&=&{1 \over N}T_{\pm\pm} \nn
&=&{1 \over 12}\left( \partial_\pm^2 \rho 
- - \partial_\pm\rho \partial_\pm\rho \right) \nn
&& +{1 \over 8} \left\{
\left( 
\partial_\pm\tilde\phi \partial_\pm\tilde\phi \right)
\rho+{1 \over 2}{\partial_\pm \over \partial_\mp}
\left( \partial_\pm\tilde\phi 
\partial_\mp\tilde\phi \right)\right\} \nn
&& +{1 \over 8}\left\{ 
- -2 \partial_\pm \rho \partial_\pm \tilde\phi 
+\partial_\pm^2 \tilde\phi \right\} + t^\pm(x^\pm) \\
\label{req2}
0&=&{1 \over N}T_{\pm\mp} \nn
&=& -{1 \over 12}\partial_+\partial_- \rho
- -{1 \over 8}\partial_+ \tilde\phi 
\partial_- \tilde \phi 
- -{1 \over 4}\partial_+\partial_-\tilde\phi \\
\label{eqtp2}
0&=&{1 \over 16}\partial_+(\rho \partial_-\tilde\phi)
+{1 \over 16}\partial_-(\rho \partial_+\tilde\phi)
- -{1 \over 8}\partial_+\partial_-\rho 
\eea
The function $t^\pm(x^\pm)$ can be absorbed into 
the choice of the coordinate and we can choose
\be
\label{tpm}
t^\pm(x^\pm)=0\ .
\ee
 Combining (\ref{eqnpp2}) and (\ref{req2}), 
we obtain
\be
\label{eqcm}
- -{1 \over 3}(\partial_\pm \rho)^2 
+{1 \over 2}\rho(\partial_\pm \tilde\phi)^2
- -\partial_\pm \rho \partial_\pm \tilde\phi =0
\ee
i.e., 
\be
\label{eqpa}
\partial_\pm \tilde\phi 
= {1 + \sqrt{1 + {2 \over 3}\rho}
\over \rho}\partial_\pm \rho \ \mbox{ or }\ 
{1 - \sqrt{1 + {2 \over 3}\rho}
\over \rho}\partial_\pm \rho \ .
\ee
This tells that
\be
\label{phi}
\tilde\phi= \int d\rho {1 \pm 
\sqrt{1 + {2 \over 3}\rho} \over \rho} \ .
\ee
 Substituting (\ref{phi}) into (\ref{eqtp2}), 
we obtain
\be
\label{ppm}
\partial_+ \partial_-\left\{
\left(1 + {2 \over 3}\rho\right)^{3 \over 2}\right\}=0
\ee
i.e.,
\be
\label{rho}
\rho={3 \over 2}\left\{-1 + 
\left(\rho^+(x^+) + \rho^-(x^-)\right)^{2 \over 3}
\right\}\ .
\ee
Here $\rho^\pm$ is an arbitrary function of 
$x^\pm = t \pm x$. 
We can straightforwardly confirm that the solutions 
(\ref{phi}) and (\ref{rho}) satisfy (\ref{req2}).
The scalar curvature is given by
\bea
\label{sR}
R&=&8\e^{-2\rho}\partial_+\partial_-\rho \nn
&=& -{8 \over 3}{\e^{-3\left\{
- -1+\left(\rho^+(x^+)+\rho^-(x^-)\right)^{2 \over 3}
\right\}} \over 
\left(\rho^+(x^+)+\rho^-(x^-)
\right)^{{4 \over 3}}}{\rho^+}'(x^+){\rho^-}'(x^-)
\eea
Note that when $\rho^+(x^+)+\rho^-(x^-)=0$, 
there is a curvature singularity. 
Especially if we choose
\be
\label{Krus}
\rho^+(x^+)={r_0 \over x^+}\ ,\ \ 
\rho^-(x^-)=-{x^- \over r_0}
\ee
there are curvature singularities at $x^+x^-=r_0^2$ and 
horizon at $x^+=0$ or $x^-=0$. The asymptotic flat 
regions are given by $x^+\rightarrow +\infty$ ($x^-<0$) 
or $x^-\rightarrow -\infty$ ($x^+>0$).
Therefore we can regard $x^\pm$ as corresponding to  
the Kruskal coordinates in 4 dimensions.

In order to discuss the Hawking radiation (which is 
usually related with trace anomaly \cite{CS}), it is 
necessary to find the exact vacuum not only at the 
classical level but even at the quantum level.
In the following, we
determine the function $\tilde\phi=\ln f(\phi)$ 
in (\ref{clac})
so that the linear dilaton vacuum 
\be
\label{ldv}
\rho=\phi=-{1 \over 2}\left(\ln x^+ + \ln x^- + \ln \lambda^2 
\right)
\ee
is an exact solution. 
 Substituting (\ref{ldv}) into (\ref{req}), we find,
\be
\label{reqd}
T_{\pm\pm}={N \over 16}\left( (\phi')^2+ \phi'' \right)
{\lambda^2 \over x^+ x^-} =0
\ee
This tells 
\be
\label{tp2}
\tilde\phi=2\ln(\phi + c)
\ee
 Substituting (\ref{tp2}) into (\ref{eqtp}) we 
find that the constant of the integration 
should vanish: $c=0$, i.e., 
\be
\label{f}
f(\phi)=\phi^2\ .
\ee
The solution (\ref{f}) when substituted to (\ref{ldv}) 
satisfies  Eq.(\ref{eqnpp}).
If we divide the energy-momentum tensor into 
classical and quantum parts, the Hawking radiation 
is given by substituting the classical solution into 
the quantum part (the part proportional to $N$).
When we substitute the shock wave solution\footnote{
The scalar field $\chi_i$ in (\ref{clac}) 
cannot make the shock wave.
Here we suppose that the dilaton gravity also 
couples with another, minimal scalar field 
which appeared in the original CGHS model \cite{12}.}
\bea
\label{sws}
\rho&=&\left\{
\begin{array}{ll}
- -{1 \over 2}\ln \left(1 + {a \over \lambda}
\e^{\lambda \sigma^-}\right) 
\hskip 1cm & \sigma < \sigma_0 \\
- -{1 \over 2}\ln \left(1 + {a \over \lambda}
\e^{\lambda (\sigma^--\sigma^+ + \sigma_0^+)} \right)
& \sigma^+ > \sigma_0 
\end{array} \right. \\
\phi&=&\left\{
\begin{array}{ll}
- -{\lambda \over 2}\sigma^+ 
- - {1 \over 2}\ln \left( \e^{-\lambda \sigma^-}
+ {a \over \lambda} \right) \hskip 1cm 
& \sigma^+ < \sigma_0 \\
- - {1 \over 2}\ln \left( {a \over \lambda}
\e^{\lambda \sigma_0}
+ \e^{\lambda(\sigma^+ - \sigma^-)} \right)  
& \sigma^+ > \sigma_0 
\end{array}
\right.
\eea
we find that  $\phi$ dependent terms 
in the quantum part of 
the energy momentum tensor vanish 
when $|\sigma^+|\rightarrow\infty$. 
This means the behavior in the asymptotic region, 
especially the Hawking radiation,  
is identical with that of the CGHS model \cite{12}.

We now investigate Bousso and Hawking's 
choice \cite{5}:
\be
\label{BHch}
f(\phi)=\e^{-2\phi}\ \  (\tilde\phi=-2\phi)\ .
\ee
Then the quantum part of the energy-momentum tensor 
has the following form:
\bea
\label{Tppe}
T^q_{\pm\pm}&=&{N \over 12}\left( \partial_\pm^2 \rho 
- - \partial_\pm\rho \partial_\pm\rho \right) \nn
&& +{N \over 2} \left\{
\left(\partial_\pm\phi \partial_\pm\phi \right)\rho
+{1 \over 2}{\partial_\pm \over \partial_\mp}
\left( \partial_\pm\phi \partial_\mp\phi 
\right)\right\} \nn
&& -{N \over 4}\left\{ 
- -2 \partial_\pm \rho \partial_\pm \phi 
+\partial_\pm^2 \phi \right\} + t(x^\pm) \\
\label{Tpm2}
T^q_{\pm\mp}&=& -{N \over 12}\partial_+\partial_- \rho
- -{N \over 2}\partial_+ \phi \partial_- \phi 
+{N \over 2}\partial_+\partial_- \phi \ .
\eea
 Substituting the classical shock wave solution
(\ref{sws}), we find when $\sigma^+ < \sigma$
\bea
\label{Tminf}
T^q_{+-}&=&{N\lambda^2 \over 8}
{1 \over \dv}  \nn
T^q_{++}&=&{N\lambda^2 \over 16}\ln\dv + t^+(\sigma^+) 
\nn
T^q_{--}&=& - {N\lambda^2 \over 48}
\left\{1 - { 1\over \dv^2}\right\} \nn
&& - {N\lambda^2 \over 16}{\ln\dv \over \dv^2}
+ {N \over 16}{a\lambda\e^{\lambda\sigma^-} \sigma^+
\over \dv^2} + t^-(\sigma^-) \ .
\eea
This tells that there is incoming energy from the 
past null infinity ($\sigma^+\rightarrow -\infty$ 
or $\sigma^-\rightarrow -\infty$). However, the explicit
estimation is  
problematic. The problem is caused by the fact that the dilaton 
vacuum is not the exact vacuum. Especially the last 
term in $T^q_{--}$, which is linear with respect to 
$\sigma^+$, tells that we cannot use the dilaton 
vacuum as a classical approximation.
The linear term also makes impossible to 
determine $t^-(\sigma^-)$ by the boundary condition 
at $\sigma^+\rightarrow -\infty$ although 
the Hawking radiation is essentially given by 
$t^-(\sigma^-)$ as we will see in the following.

When $\sigma^+ > \sigma_0$, we find 
\bea
\label{Tmsw}
T^q_{+-} &=& {N\lambda^2 \over 12}{1 \over \ds^2}
- -{N\lambda^2 \over 6}{1 \over \ds} \nn
T^q_{\pm\pm}
&=&-{N\lambda^2 \over 48} \left\{ 1 + {1 \over \ds^2}
\right\} \nn
&&-{N\lambda^2 \over 16}{\ln \ds - 1 \over \ds^2}
+t^\pm (\sigma^\pm) 
\eea
Then when $\sigma^+\rightarrow +\infty$, the energy 
momentum tensor behaves as
\bea
\label{asT}
T^q_{+-}&\rightarrow& -{N\lambda^2 \over 12}\ ,\nn
T^q_{\pm\pm}&\rightarrow& {N\lambda^2 \over 48} 
+ t^\pm(\sigma^\pm)
\eea
This expresses the Hawking radiation but we cannot 
determine the unknown function $t^-(\sigma^-)$.

Hence, we found that there maybe new contributions to Hawking 
radiation from the dilaton dependent terms in the trace 
anomaly. However, in order to make their accurate estimation 
one has to construct new solvable models of 2D black holes 
with an arbitrary $f$ and (or) another choices for 
dilatonic couplings $Z$, $C$ and $V$ 
in general model of dilatonic gravity.

Finally in this section, we evaluate the 
contribution to the black hole entropy from 
$W$ in eq.(\ref{qc}). 
The contribution from the first classical term maybe 
investigated by standard methods \cite{SB}.
Following this procedure ,  
the contribution from the dilaton dependent terms
\be
\label{other}
- -{1 \over 2}\int d^2x \sqrt{-g} \Bigl[ 
- -{N \over 16\pi}{{f'}^2 \over f^2}
(\nabla^\lambda \phi)
(\nabla_\lambda \phi) {1 \over \Delta}R 
- -{N \over 8\pi}\ln f R \Bigr]\ .
\ee
can be evaluated as follows
\be
\label{ent}
- -{N \over 16}\int d^2x \sqrt{-g} \left(
{{f'}^2 \over f^2}(\nabla^\lambda \phi)
(\nabla_\lambda \phi) \psi \right) 
- -{N \over 4\pi}\ln f(\phi_0)\ .
\ee
Here $\phi_0$ is the value of the classical 
solution for the dilaton field at the horizon:
\be
\label{phi0}
\phi_0=-{1 \over 2}\ln \left(M \over \lambda\right)
\ee
and $\psi$ is defined by
\be
\label{psi} 
\Delta \psi = \delta(r)
\ee
with the boundary condition where
\be
\label{bc}
\psi\rightarrow \ln r \ ,\ \ \mbox{when}\ r\rightarrow 0
\ .
\ee
Here we choose the coordinate system where the metric 
of the black hole is given by 
\be
\label{rco}
ds^2=dr^2 + \sinh^2 \sqrt{M \over \lambda} dt^2\ .
\ee
when Wick-rotated to the Euclidean signature. Hence, at least 
on qualitative level we see the appearence of new terms in 
quantum corrections to black hole entropy.

\section{Trace anomaly and induced effective action 
for 4D dilaton coupled scalar}

It could be interesting to generalize the results of second section
 for 4D case. The purpose of the present section will be 
to calculate the non-local effective action for 4D dilaton coupled 
conformal scalar.
Let us consider the theory with the following 
Lagrangian in curved spacetime (we work in Minkowski signature)
\be
\label{v1}
L=\varphi f(\phi) (\Box - \xi R )\varphi
\ee
where $\varphi$ is quantum scalar field, 
$\Box=g^{\mu\nu}\nabla_\mu \nabla_\nu$, 
$\phi$ is an external field (dilaton), 
$f(\phi)$ is an arbitrary function.

It is very easy to check that for 
conformal transformation 
\bea
\label{v2}
g_{\mu\nu}&\rightarrow & \e^{2\sigma}g_{\mu\nu} \ ,\nn
R&\rightarrow & \e^{-2\sigma}
\left(R-6\Box\sigma - 
6 (\nabla_\mu\sigma )(\nabla^\mu\sigma)\right)
\eea
the theory with Lagrangian (\ref{v1}) is conformally 
invariant for $\xi={1 \over 6}$.

Let us calculate the divergent part of the effective 
action for the theory (\ref{v1}):
\bea
\label{v3}
\Gamma_{\rm div}&=&-{i \over 2}\Tr\ln
\left\{f(\phi)\left[ \Box - \xi R 
+{\Box f(\phi) \over 2f(\phi)}
+{(\nabla^\mu f(\phi) ) \over f(\phi)}\nabla_\mu 
\right]\right\} \nn
&=& {1 \over (n-4)}\int d^4x \sqrt{-g} 
b_4
\eea
where $b_4$ is the $b_4$-coefficient of 
Schwinger-De Witt expansion.
The methods of its calculation are well known 
(see, for example, section 3.6 of ref.\cite{3}).
Applying these methods, after some algebra we will 
get 
\bea
\label{v4}
b_4 &=& {1 \over 2}\left({1 \over 6} - \xi\right)^2 R^2 
+ {1 \over 4} {(\nabla f)^2 \over f^2}
\left({1 \over 6} - \xi\right)R 
+{1 \over 32}{[(\nabla f)( \nabla f)]^2 \over f^4} \nn
&& +{1 \over 2}\left({1 \over 6} - \xi\right)\Box R 
+{1 \over 24}\Box \left(
{(\nabla f) (\nabla f) \over f^2}\right) \nn
&& + {1 \over 180}
\left( R_{\mu\nu\alpha\beta}^2 - R_{\mu\nu}^2 
+ \Box R \right)
\eea
Note that $f(\phi)$-multiplier in Eq.(\ref{v3}) 
does not give the contribution to $b_4$.

For $\xi = {1 \over 6}$, we get the trace anomaly:
\be
\label{v5}
<T_\mu^{\, \mu}>=b_4
\ee
Hence the trace anomaly for dilaton 
coupled 4$D$ scalar is given by 
\bea
\label{v6}
T&=&{1 \over (4\pi)^2}
\Bigl\{ {1 \over 32}{[(\nabla f) (\nabla f)]^2 \over f^4}
+{1 \over 24}\Box \left(
{(\nabla f) (\nabla f) \over f^2}\right) \nn
&& + {1 \over 180}
\left( R_{\mu\nu\alpha\beta}^2 - R_{\mu\nu}^2 
+ \Box R \right)\Bigr\}
\eea
Here, the last term is the well-known conformal 
anomaly (for a review, see \cite{MD}) for conformally invariant scalar. 
The first two terms in (\ref{v6}) are the 
dilaton contribution to conformal anomaly.

Let us write the Eq.(\ref{v6}) in a slightly 
different form:
\bea
\label{v7}
T&=&\Bigl\{b\left(F+ {2 \over 3}\Box R \right) 
+ b'G + b'' \Box R +  \nn
&& a_1 
{[(\nabla f) (\nabla f)]^2 \over f^4}
+a_2\Box \left(
{(\nabla f) (\nabla f) \over f^2}\right) \Bigr\}
\eea
where $F$ is the square of Weyl tensor in four 
dimensions, $G$ is Gauss-Bonnet invariant.
For scalar field, it follows from (\ref{v6}) that 
\be
\label{v8}
b={ 1 \over 120 (4\pi)^2 }\ ,\ \ 
b'=-{ 1 \over 360 (4\pi)^2 }\ ,\ \ 
a_1={ 1 \over 32 (4\pi)^2 }\ ,\ \ 
a_2={ 1 \over 24 (4\pi)^2 }
\ee
and in principle $b''$ is an arbitrary parameter 
(it maybe changed by the finite renormalization 
of local counterterm).

The non-local effective action induced by the 
conformal anomaly (without dilaton) has been 
calculated sometime ago \cite{R}.
Using the equation 
\be
\label{v9}
T={1 \over \sqrt{-g}}{\delta \over \delta \sigma}
W(\sigma)
\ee
and integrating it, one can restore the non-local 
effective action $W$ induced by the conformal 
anomaly:
\bea
\label{v10}
W&=&b\int d^4x \sqrt{-g} F\sigma \nn
&& +b'\int d^4x \sqrt{-g} \Bigl\{\sigma\left[
2\Box^2 + 4 R^{\mu\nu}\nabla_\mu\nabla_\nu 
- - {4 \over 3}R\Box + {2 \over 3}(\nabla^\mu R)\nabla_\mu 
\right]\sigma \nn
&& + \left(G-{2 \over 3}R\right)\sigma \Bigr\} \nn
&& -{1 \over 12}\left(b'' + {2 \over 3}(b + b')\right)
\int d^4x \sqrt{-g}\left[R - 6 \Box \sigma 
- - 6(\nabla \sigma)(\nabla \sigma) \right]^2 \nn
&& + \int d^4x \sqrt{-g} \Bigl\{ 
a_1 {[(\nabla f) (\nabla f)]^2 \over f^4}\sigma 
+a_2\Box \left({(\nabla f) (\nabla f) \over f^2} 
\right)\sigma \nn
&& \hskip 2cm 
+a_2{(\nabla f) (\nabla f) \over f^2} [(\nabla \sigma) 
(\nabla \sigma )]\Bigr\}
\eea
Here the $\sigma$-independent term is dropped, last terms 
represent the contribution from the dilaton dependent 
 terms in trace anomaly. 
Similarly one can calculate trace anomaly and induced 
effective action for other theories like dilaton 
coupled spinor or dilaton coupled Weyl gravity with 
Lagrangian:
\be
\label{v11}
L=F(\phi)C_{\mu\nu\alpha\beta}C^{\mu\nu\alpha\beta}
\ee
or dilaton coupled vector: 
$L=-{1 \over 4}g(\phi)F_{\mu\nu}F^{\mu\nu}$. 
Notice that in 2D case such vector field is not conformally 
invariant one.

\section{Conformal sector of dilaton gravity and 
quantum cosmology}

Let us consider now the classical theory of 
dilatonic gravity:
\be
\label{v12}
L_{cl}=Z(\phi)g^{\mu\nu}\partial_\mu\phi 
\partial_\nu \phi + C(\phi)R + V(\phi)
\ee
where $Z$, $C$, $V$ are the arbitrary 
dilatonic functions.
For specific choice of these functions the 
theory (\ref{v12}) represents the low-energy 
string effective action or Brans-Dicke gravity. 
So it may be considered as string-motivated classical 
gravity.

Adding the induced action to the action (\ref{v12})
(where part of linear on $\sigma$-terms are dropped 
away), we get in the case of conformally flat 
fiducial metric $g_{\mu\nu}=\e^{2\sigma}\eta_{\mu\nu}$:
\bea
\label{v13} 
S&=&W + S_{cl} \nn
&=&\int d^4x \Bigl\{
2b'(\Box \sigma )^2 
- -\left[3b'' + 2(b + b')\right]\left[
\Box \sigma + (\partial_\mu \sigma)^2\right]^2 \nn
&& +a_1 {[(\nabla f) (\nabla f)]^2 \over f^4}\sigma 
+a_2\Box \left({(\nabla f) (\nabla f) \over f^2} \right) 
\sigma \nn
&& +a_2{(\nabla f) (\nabla f) \over f^2} g^{\mu\nu}
(\nabla_\mu \sigma) (\nabla_\nu \sigma) \nn
&&+ \e^{2\sigma}Z(\phi)g^{\mu\nu}\partial_\mu\phi 
\partial_\nu \phi 
+ \e^{2\sigma}C(\phi)\left[
- -6\Box\sigma - 6 (\partial_\mu\sigma )
(\partial^\mu\sigma)\right] \nn
&& + \e^{4\sigma}V(\phi) \Bigr\}
\eea
The action (\ref{v13}) describes the conformal 
sector of $4D$ dilatonic gravity. 
It is direct generalization of the conformal sector 
of $4D$ gravity which was introduced and studied 
in refs.\cite{AMO}.

That is very interesting problem for future research to study the quantum structure of the theory with the 
action (\ref{v13}), its properties, the existence 
of fixed points, etc. 
In particulary, one can expect as it happened with its 
analog for $\phi={\rm const}$ \cite{AMO} that is may 
provide the solution of the cosmological constant 
problem. The gravitational dressing of matter beta-functions 
in such theory may lead to the interesting consequences for 
Standard Model and GUTs \cite{PO}.

The classical solutions of the theory (\ref{v13}) 
should define the cosmology of early universe 
with back-reaction of the conformal 
dilaton coupled matter.
However, it is not easy to search for solutions of the 
theory (\ref{v13}). (Of course, one can work again in 
large-N expansion what justifies the neglecting of 
classical term in (\ref{v13})). 
So we will start from dilaton coupled Weyl gravity as 
the classical gravity. Adding to the action of such theory 
the induced effective action we omit the linear on $\sigma$
terms.
(That maybe justified by adding to the theory of 
dilaton and gravity dependent source for $\sigma$).
Working on conformally flat 
metric $g_{\mu\nu}=\e^{2\sigma}\eta_{\mu\nu}$ (where 
classical action (\ref{v11}) is equal to zero), we 
may find the following classical solutions:
\be
\label{v14}
\sigma=\alpha \ln H_1 \eta\ ,\ \ \ 
f=\beta \ln H_2\eta
\ee
where $H_1$, $H_2$, $\alpha$ and $\beta$ are 
some constants.
Their explicit values are defined by the complicated 
algebraic system of two equations
\be
\label{v15}
{\delta W \over \delta \sigma}=0\ ,\ \ 
{\delta W \over \delta f}=0\ .
\ee
Note that for the same $a_1$, $a_2$-coefficients 
in $W$ (\ref{v10}) one can change the coefficients 
$b$, $b'$ and $b''$ by adding the conformal matter 
minimally interacting with the dilaton.
Hence the solutions (\ref{v14}) define the whole 
class of metrics.
In particular, for $\alpha=-1$, we get the solution 
which corresponds to the inflationary universe of Starobinsky type  
\cite{S}, however now with non-trivial dilaton. 
One can investigate other types of solutions for induced 
effective action, for example, black hole type solutions.

\section{Summary}

In summary, trace anomaly and induced action for dilaton coupled 
scalar in 2D and 4D dimensions are found. The large-$N$ 
effective action for quantum dilaton-scalar gravity is also 
evaluated. The appearence of 
new, dilaton dependent terms in the effective action is shown. 
Some preliminary results on the role of these terms for 2D black 
holes and Hawking radiation 
are reported. The conformal sector of 4D dilatonic gravity 
is constructed 
and quantum cosmology is discussed. 

Our results bring to the attention a number of problems. 
Let us mention some of them:
\begin{enumerate}
\item The construction of classically solvable dilaton gravities 
with dilaton 
coupled scalars. Search for new black holes in such models. 
Calculation 
of new corrections to Hawking radiation and black hole entropy.
\item  Trace anomaly for 4D dilaton coupled vector, 
spinor and graviton. Study of 
quantum cosmology with back reaction of such fields.
\item Study of one-loop renormalizability of the theory (\ref{v11}).
\item Investigation of quantum structure for conformal sector of
 dilatonic gravity.
\item Generalizations of C-theorem with account of dilaton 
dependent terms.
\end{enumerate}
We hope to return to the study of some of these questions in near 
future.

\ 

\noindent
{\bf Acknoweledgments} We would like to thank R. Bousso 
and S. Hawking for 
email discussions. We are grateful to T. Muta and whole 
Particle Theory 
Group of Hiroshima University for kind hospitality while 
completing this 
work. Our research has been partly supported by COLCIENCIAS 
(Colombia) 
and JSPS (Japan).

\end{document}